\documentstyle[psfig,twocolumn,prl,aps,epsf,epsfig]{revtex}

\newcommand{\hept}{$^4$He$(\vec{e},e^\prime\vec{p}\,)\,^3$H}

\begin{document}

\title{Polarization Transfer in the $^4$He($\vec e,e^\prime\vec p\,$)$^3$H
Reaction up to $Q^2$ = 2.6 (GeV/c)$^2$}

\author {
S.~Strauch,$^{1}${\cite{atgwu}}
S.~Dieterich,$^{1}$
K.A.~Aniol,$^{2}$
J.R.M.~Annand,$^{3}$
O.K.~Baker,$^{4,5}$
W.~Bertozzi,$^{6}$
M.~Boswell,$^{7}$
E.J.~Brash,$^{8}$
Z.~Chai,$^{6}$
J.-P.~Chen,$^{5}$
M.E.~Christy,$^{4}$
E.~Chudakov,$^{5}$
A.~Cochran,$^{4}$
R.~De Leo,$^{9}$
R.~Ent,$^{5}$
M.B.~Epstein,$^{2}$
J.M.~Finn,$^{10}$
K.G.~Fissum,$^{11}$
T.A.~Forest,$^{12}$
S.~Frullani,$^{13}$
F.~Garibaldi,$^{13}$
A.~Gasparian,$^{4}$
O.~Gayou,$^{10,14}$
S.~Gilad,$^{6}$
R.~Gilman,$^{1,5}$
C.~Glashausser,$^{1}$
J.~Gomez,$^{5}$
V.~Gorbenko,$^{15}$
P.L.J.~Gueye,$^{4}$
J.O.~Hansen,$^{5}$
D.W.~Higinbotham,$^{6}$
B.~Hu,$^{4}$
C.E.~Hyde-Wright,$^{12}$
D.G.~Ireland,$^{3}$
C.~Jackson,$^{4}$
C.W.~de~Jager,$^{5}$
X.~Jiang,$^{1}$
C.~Jones,$^{4}$
M.K.~Jones,$^{16}$
J.D.~Kellie,$^{3}$
J.J.~Kelly,$^{16}$
C.E.~Keppel,$^{4}$
G.~Kumbartzki,$^{1}$
M.~Kuss,$^{5}$
J.J.~LeRose,$^{5}$
K.~Livingston,$^{3}$
N.~Liyanage,$^{5}$
R.W.~Lourie,$^{17}$
S.~Malov,$^{1}$
D.J.~Margaziotis,$^{2}$
D.~Meekins,$^{18}$
R.~Michaels,$^{5}$
J.H.~Mitchell,$^{5}$
S.K.~Nanda,$^{5}$
J.~Nappa,$^{1}$
C.F.~Perdrisat,$^{10}$
V.A.~Punjabi,$^{19}$
R.D.~Ransome,$^{1}$
R.~Roch\'e,$^{18}$
G.~Rosner,$^{3}$
M.~Rvachev,$^{6}$
F.~Sabatie,$^{12}$
A.~Saha,$^{5}$
A.~Sarty,$^{18}$
J.M.~Udias,$^{20}$
P.E.~Ulmer,$^{12}$
G.M.~Urciuoli,$^{13}$
J.F.J.~van~den~Brand,$^{21}$
J.R.~Vignote,$^{20}$
D.P.~Watts,$^{3}$
L.B.~Weinstein,$^{12}$
K.~Wijesooriya,$^{22}$
and B.~Wojtsekhowski$^{5}$
}

\address{
$^{1}${Rutgers, The State University of New Jersey, Piscataway, New Jersey 08854} \\
$^{2}${California State University, Los Angeles, California 90032} \\
$^{3}${University of Glasgow, Glasgow, G12 8QQ, Scotland, United Kingdom} \\
$^{4}${Hampton University, Hampton, Virginia 23668} \\
$^{5}${Thomas Jefferson National Accelerator Facility, Newport News, Virginia 23606} \\
$^{6}${Massachusetts Institute of Technology, Cambridge, Massachusetts 02139} \\
$^{7}${Randolph-Macon Woman's College, Lynchburg, Virginia 24503} \\
$^{8}${University of Regina, Regina, Saskatchewan, Canada S4S 0A2} \\
$^{9}${INFN, Sezione di Bari and University of Bari, I-70126, Bari, Italy} \\
$^{10}${College of William and Mary, Williamsburg, Virginia 23187} \\
$^{11}${University of Lund, SE-221 00 Lund, Sweden} \\
$^{12}${Old Dominion University, Norfolk, Virginia 23529} \\
$^{13}${INFN, Sezione Sanit\'a and Istituto Superiore di Sanit\'a, Laboratorio di Fisica, I-00161 Rome, Italy} \\
$^{14}${Universit\'e Blaise Pascal, F-63177 Aubi\`ere, France} \\
$^{15}${Kharkov Institute of Physics and Technology, Kharkov 310108, Ukraine} \\
$^{16}${University of Maryland, College Park, Maryland 20742} \\
$^{17}${State University of New York at Stony Brook, Stony Brook, New York 11794} \\
$^{18}${Florida State University, Tallahassee, Florida 32306} \\
$^{19}${Norfolk State University, Norfolk, Virginia 23504} \\
$^{20}${Universidad Complutense de Madrid, E-28040 Madrid, Spain} \\
$^{21}${Vrije Universiteit, NL-1081 HV Amsterdam, Netherlands} \\
$^{22}${University of Illinois at Urbana-Champaign, Urbana, Illinois 61801} \\
}

\maketitle

\begin{abstract}
We have measured the proton recoil polarization in the $^4$He($\vec e,e^\prime
\vec p\,$)$^3$H reaction at $Q^2$ = 0.5, 1.0, 1.6, and 2.6 (GeV/c)$^2$.
The measured ratio of polarization transfer coefficients differs from a fully
relativistic calculation, favoring the inclusion of a predicted medium
modification of the proton form factors based on a quark-meson coupling model.
In contrast, the measured induced polarizations
agree reasonably well with the fully relativistic calculation indicating
that the treatment of final-state interactions is under control.
\end{abstract}

\date{\today}

The underlying theory of strong interactions is Quantum ChromoDynamics (QCD),
yet there are no ab-initio calculations of nuclei available. Nuclei are
effectively and well described as clusters of protons and neutrons held
together by a strong, long-range force mediated by meson exchange,
whereas the saturation properties of nuclear matter arise from the
short-range, repulsive part of the strong interaction \cite{Mosz}.
Whether the nucleon bound in the nuclear medium changes structure
has been a long-standing issue in nuclear physics.
At nuclear densities of about 0.17 fm$^{-3}$ nucleon wave functions
have significant overlap.
In the chiral limit, one expects nucleons to lose their identity
altogether and nuclei to make a transition to a quark-gluon plasma.

Unfortunately, distinguishing possible changes in the structure of nucleons
embedded in a nucleus from more conventional many-body effects is only
possible within the context of a model. Nucleon modifications can be described
in terms of coupling to excited states, and such changes are
intrinsically intertwined with many-body effects, such as meson-exchange
currents (MEC) and isobar configurations (IC).
Therefore, interpretation of an experimental signature as an indication of
modifications of the nucleon form factors only makes sense if this
results in a more economical effective description of the bound, quantum,
nuclear many-body system.

The quark-meson coupling (QMC) model of Lu {\sl et al.}~\cite{Lu98}
suggests a measurable deviation of the ratio of the proton's electric ($G_E$)
and magnetic ($G_M$) form factors from its free space value over the
$Q^2$ range accessible by experiment. This calculation is consistent with
present constraints on possible medium modifications for both $G_E$ 
(from the Coulomb Sum Rule, with $Q^2$ $<$ 0.5 (GeV/c)$^2$ \cite{Jo95,Mo01,Ca02}),
$G_M$ (from a $y$-scaling analysis \cite{Si88}, for $Q^2$ $>$ 1 (GeV/c)$^2$),
and limits on the scaling of nucleon magnetic moments in 
nuclei \cite{ericrich}. Similar effects have been calculated in the 
light-front constituent quark model of Frank {\sl et al.}~\cite{Fr96}.

In unpolarized A($e,e^\prime p$) experiments involving light- and
medium-heavy nuclei, deviations were observed in the longitudinal/transverse
character of the nuclear response compared to the free proton case
\cite{steen,ulmer,reffay}. Below the two-nucleon emission threshold,
these deviations were originally interpreted
as changes in the nucleon form factors within the nuclear medium.
However, strong interaction effects on the ejected proton
(final state interactions [FSI]) later also succeeded in explaining the
observed effect \cite{cohen}. 
This illustrates that any interpretation in terms of medium modifications
to nucleon form factors requires having excellent control of FSI effects.

For free electron-nucleon scattering, the ratio of the electric to
magnetic Sachs form factors, ($G_E/G_M$), is directly proportional to
the ratio of the transverse and longitudinal transferred
polarizations, ($P'_x/P'_z$) \cite{Ar81,Dipc}.  This relationship was
recently used to extract $G_E/G_M$ for the proton
\cite{gegmprl,gegmprc,gayouprl}. Polarization transfer in quasielastic nucleon
knockout remains sensitive to this ratio of form factors (possibly modified
by the nuclear medium).
A variety of calculations for the A($\vec e,e^\prime \vec p\,$)
reaction indicate that FSI and MEC effects on polarization transfer
observables are small, amounting to only a $<$~10\%
correction \cite{laget,Ud98,kelly}. In addition, these nuclear interaction
effects tend to largely cancel in the ratio of polarization transfer
coefficients $P'_x/P'_z$. 

Recently, polarization transfer for the \hept{} reaction
at $Q^2$ = 0.4 (GeV/c)$^2$ was studied \cite{mainz4he}. The addition of
medium-modified proton form factors, as predicted by the QMC model,
to a state-of-the-art fully relativistic model \cite{Ud98} gave a
good description of the data. The authors concluded that, within the
model space examined, the data favor
models with medium-modified form factors over those with free form factors,
but the latter could not be excluded. Examination of this finding
over a larger range in $Q^2$ seems an obvious step for further investigation.

The experiment reported here includes measurements of the polarization transfer
coefficients over the range of $Q^2$ from 0.5 to 2.6 (GeV/c)$^2$, and
as a function of missing momentum in the range 0 to 240 MeV/c, in
order to maximize sensitivity to the electric to magnetic form factor
ratio for protons bound in the $^4$He nucleus. This nucleus was selected
for study because its relative simplicity allows realistic microscopic
calculations and its high density enhances any possible medium effects. 
As the experiment was designed to detect differences between the
in-medium polarizations and the free values, both $^4$He and
$^1$H targets were employed (except at $Q^2=2.6$ (GeV/c)$^2$, where only
$^4$He data were acquired due to beam time constraints). 

Kinematics settings for the present experiment in Hall A at Jefferson Lab
(JLab) are given in Table I.  The experiment used beam currents of 40 $\mu$A
for the lower $Q^2$ values and up to 70 $\mu$A for the highest $Q^2$
value, combined with beam polarizations of 66\% for the lowest $Q^2$
value and $\approx$ 77\% for the other $Q^2$ values. The beam helicity
was flipped pseudorandomly to reduce systematic errors of the
extracted polarization transfer observables.  The proton spectrometer
was equipped with a focal plane polarimeter (FPP)
\cite{markfpp,nimfpp}.  Polarized protons lead to azimuthal
asymmetries after scattering in the carbon analyzer of the FPP. These
distributions, in combination with information on the beam helicity,
were analyzed by means of a maximum likelihood method to obtain the
induced and transferred polarization components. More details on the
analysis can be found in Refs. \cite{gegmprl,malov,sonjaphd}.

\begin{figure}
\centerline{\epsfig{file=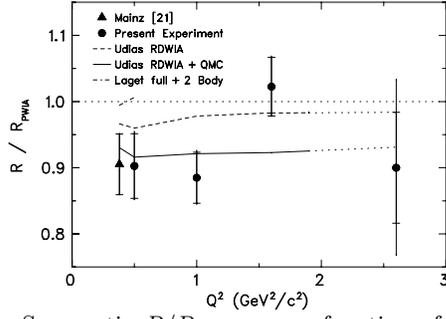,width=2.30in,angle=90}}
\caption{Superratio $R$/$R_{PWIA}$ as a function of $Q^2$.
$R$ is defined as the double ratio
($P^\prime_x/P^\prime_z$)$_{He}$/($P^\prime_x/P^\prime_z$)$_{H}$.
In PWIA (short-dashed curve) this superratio is identically unity, barring
acceptance-averaging effects. The dashed
curve shows the results of the full relativistic calculation of 
Udias {\it et al.} \protect\cite{Ud98}. The dot-dashed curve shows the
results of Laget's full calculation, including two-body currents
\protect\cite{laget}. The solid curve indicates the full relativistic
calculation of Udias including medium modifications
as predicted by a quark-meson coupling model \protect\cite{Lu98}.
For $Q^2 >$ 1.8 (GeV/c)$^2$ the Udias calculations maintain a constant
relativistic optical potential and are indicated as short-dashed curves.
Lines connect the acceptance-averaged theory calculations and
are to guide the eye only.
\label{fig:ratioplot}}
\end{figure}

Our results are shown in Fig.~\ref{fig:ratioplot} as $R/R_{PWIA}$
for all four values of $Q^2$. $R_{PWIA}$ is the prediction based on
the relativistic plane-wave impulse approximation (RPWIA) calculation.
Here, $R$ is defined as
\begin{equation}
R = \frac{(P_x'/P_z')_{^4\rm He}}{(P_x'/P_z')_{^1\rm H}}
\label{eq:rexp}
\end{equation}
for the data, whereas $R_{PWIA}$ is the same ratio based on
the relativistic plane-wave impulse approximation (RPWIA) calculation.
The helium polarization ratio is normalized to the hydrogen
polarization ratio measured at the same setting. Such a polarization double
ratio nearly cancels all systematic uncertainties. As a cross
check, the hydrogen results were also used to extract the free proton
form factor ratio $G_E/G_M$ and found to be in excellent agreement
with previous data \cite{gegmprl,gegmprc}.
In addition, our result at $Q^2$ = 0.5 (GeV/c)$^2$ closely coincides
with the recent results at $Q^2$ = 0.4 (GeV/c)$^2$ of Mainz \cite{mainz4he},
also shown in Fig.~\ref{fig:ratioplot}.
Our experimental results for helium and hydrogen separately, in terms of
$(P_x'/P_z')$, are tabulated in Table II. Systematic uncertainties
are mainly due to possible minor misalignments of the magnetic elements
of the proton spectrometer and uncertainties in the spin
transport through these magnetic elements. They are estimated to contribute
less than 1.7\% to $R$.

The theoretical calculations by the Madrid group \cite{Ud98} are averaged
over the experimental acceptance. We note that these
relativistic calculations provide good descriptions of, {\it e.g.}, the induced
polarizations measured at Bates in the $^{12}$C($e,e^\prime \vec p\,$)
reaction \cite{batesfpp} and of $A_{TL}$ in 
$^{16}$O($e,e^\prime p\,$) as previously measured at JLab \cite{alt}.

At $Q^2$ = 0.5 and 1.0 (GeV/c)$^2$ the RPWIA
calculation overestimates the data by $\approx$~10\%.
The relativistic distorted-wave impulse approximation (RDWIA) calculation
gives a slightly smaller ($\approx$ 3\%) value of $R$ but still
overpredicts the data. After including the (density-dependent)
medium-modified form factors
as predicted by Lu {\it et al.} \cite{Lu98} in the RDWIA calculation,
excellent agreement is obtained at both settings. All calculations
shown use the Coulomb gauge, the $cc1$ current operator as defined
in \cite{Fo83}, and the MRW optical potential of \cite{Mc83}.
The $cc2$ current operator gives slightly higher values of $R$,
worsening agreement with the data. In general, various choices for, {\it e.g.},
spinor distortions, current operators, and relativistic corrections,
affect the theoretical predictions by $\le$3\%, and can presently not explain
the disagreement between the data and the RDWIA calculations.
In contrast, the datum at $Q^2$ = 1.6 (GeV/c)$^2$ is well described by the
RPWIA and RDWIA calculations, whereas all calculations are consistent with
the datum at $Q^2$ = 2.6 (GeV/c)$^2$.

A statistical analysis of the measured double ratios, including the result of
the Mainz experiment \cite{mainz4he}, and various theoretical
predictions was performed. The model space we examined encompassed the
RPWIA and RDWIA calculations of Udias {\it et al.} \cite{Ud98}, the 
latter with and without medium modifications as predicted by a
quark-meson coupling model \cite{Lu98}, the full nonrelativistic
model of Debruyne {\it et al.} \cite{Ry99,debruynephd}, and the
full nonrelativistic calculation of Laget
including two-body currents \cite{laget}. For the latter calculation
only data up to $Q^2$ = 0.5 (GeV/c)$^2$ are taken into account.
A significantly better description is given by the RDWIA calculation
when medium modifications are included.

\begin{figure}
\centerline{\epsfig{file=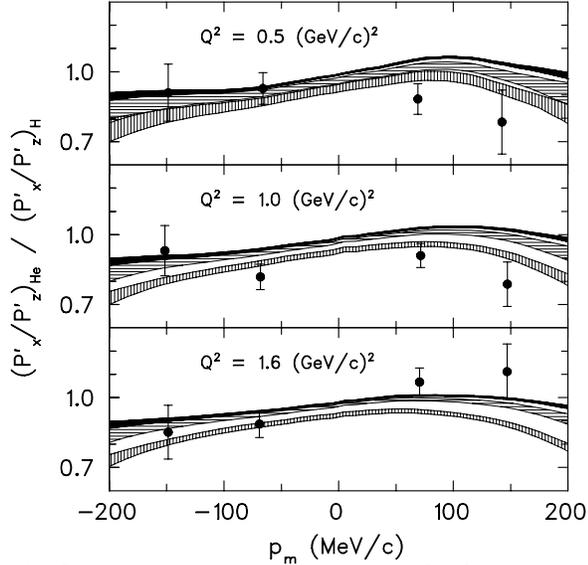,width=3.0in}}
\caption{Measured values of the polarization double ratio $R$ for 
$^4$He$(\vec e,e'\vec p\,)^3$H at $Q^2$ = 0.5~(GeV/c)$^2$ (top),
$Q^2$ = 1.0~(GeV/c)$^2$ (middle), and $Q^2$ = 1.6~(GeV/c)$^2$ (bottom).
The shaded bands represent RPWIA calculations (solid),  relativistic DWIA
calculations (horizontal dashes) and relativistic DWIA calculations
including QMC medium--modified form factors \protect\cite{Lu98} by
Udias {\it et al.} \protect\cite{Ud98} (vertical dashes). 
The bands reflect variations due to choice of current operator,
optical potential, and bound-state wave function
(see also Ref. \protect\cite{mainz4he}).
\label{fig:results}}
\end{figure}

Figure~\ref{fig:results} shows the polarization double ratio $R$ as a function
of missing momentum for the lower three $Q^2$ kinematics (the statistics
at the $Q^2$ = 2.6 (GeV/c)$^2$ kinematics are not sufficient to make a
meaningful comparison with calculations). Negative values of missing momentum
correspond to the recoiling nuclei having a momentum component
antiparallel to the direction of the three-momentum transfer.  Both
the RPWIA and the RDWIA give a reasonable, but not
perfect, description of the missing momentum dependence of the
data. As already seen in Fig.~\ref{fig:ratioplot}, the difference in magnitude
between the RDWIA calculation and the data at $Q^2$ = 0.5 and 1.0 (GeV/c)$^2$
can be largely eliminated by including the QMC medium modifications, whereas
at $Q^2$ = 1.6 (GeV/c)$^2$ the calculation without QMC medium modifications
already gives a satisfactory description. 
More precise data could unambiguously settle whether this is just a
statistical fluctuation, and would constitute a demanding test of modern
nucleon-meson descriptions of nuclear physics.

\begin{figure}
\centerline{\epsfig{file=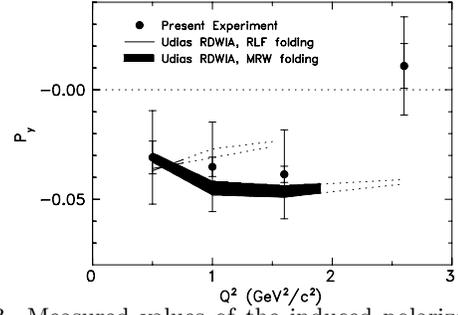,width=2.3in,angle=90}}
\caption{Measured values of the induced polarizations for the 
{$^4$He$(e,e^\prime\vec{p}\,)\,^3$H} reaction. The inner uncertainty
is statistical only; the total uncertainty includes a systematic uncertainty
of $\pm$0.02, due to imperfect knowledge of the false asymmetries.
The solid and dashed curves show the results for the full relativistic
RDWIA calculations of Udias {\it et al.} \protect\cite{Ud98}, using
differing relativistic optical potentials \protect\cite{Mc83,rlf}.
For the dashed curves, variation within the chosen
optical potential parameters is indicated by the shaded area.
The short-dashed lines indicate the $Q^2$ regions beyond the validity
of the relativistic optical potentials used.
\label{fig:induced}}
\end{figure}

Lastly, we show in Fig.~\ref{fig:induced} the induced polarization, $P_y$,
obtained by properly averaging over
the two beam helicities, and corrected for (small) false asymmetries,
as a function of $Q^2$. $P_y$ is identically zero in the absence of FSI effects
(in the one-photon exchange approximation) and constitutes a stringent test
of the validity of the inclusion of FSI effects in the calculations.
For example, an underestimate of reaction mechanism effects in the present
calculation may be due to the neglect of the
charge exchange ($\vec e$,e$^\prime \vec n\,$)($\vec n,\vec p\,$) reaction
in the RDWIA calculations. However, the measured induced polarizations agree
well with the RDWIA calculations. In addition, the
$^{12}$C($\vec e,e^\prime \vec p\,$) and $^{16}$O($\vec e,e^\prime \vec p\,$)
reactions were calculated to be insensitive to this effect \cite{kelly}.

One sees in Fig.~\ref{fig:induced} that the induced polarizations are
small for all measured $Q^2$ values.  The dashed and dot-dashed curves
represent RDWIA calculations by Udias {\it et al.}
\protect\cite{Ud98} with the MRW \cite{Mc83} and RLF \cite{rlf} relativistic
optical potentials. For the induced polarization case, the RDWIA
curves with and without medium modifications are identical: as
mentioned earlier the QMC model incorporates modifications only to the
one-body form factors.  For a rigorous calculation of the
$^4$He($e,e^\prime\vec p$)$^3$H results presented here, one would need
to take into account possible medium modifications to both one-body
form factors and many-body FSI effects.  Figure \ref{fig:induced}
confirms the expected small values of the induced polarizations, and
indicates reasonable agreement with the RDWIA calculations.

In summary, we have measured recoil polarization in
the \hept{} reaction in the range from 
$Q^2$ = 0.5 to 2.6 (GeV/c)$^2$. The datum at the lowest $Q^2$ agrees well
with the results of a recently reported Mainz measurement \cite{mainz4he}.
Such polarization transfer data are calculated to be only slightly
dependent ($<$~10\% effect) on nuclear structure effects and fine details of
the reaction mechanism. Furthermore, these effects tend to cancel in the
$P'_x/P'_z$ polarization transfer ratio.
Within our model assumptions we find strong evidence for a medium modification;
a calculation incorporating a predicted medium modification based on the
quark-meson coupling model \cite{Lu98} gives a good description of our data.
Moreover, the calculated induced polarizations agree well with our
data, giving credibility to the validity of the treatment of FSI effects
in the model.
These data provide the most stringent test to date of the
applicability of conventional meson-nucleon calculations.

\medskip

The collaboration wishes to acknowledge the Hall A technical staff and the
Jefferson Lab Accelerator Division for their outstanding support.
The Southeastern Universities Research Association (SURA) operates the Thomas
Jefferson National Accelerator Facility for the United States Department of
Energy under contract DE-AC05-84ER40150.
This work was supported by research grants from the United States Department
of Energy and the National Science Foundation, the Italian Istituto di Fisica
Nucleare (INFN), the Natural Sciences and Engineering Council of Canada (NSERC),
the Swedish Natural Science Research Council,
and the Comunidad de Madrid and Ministerio de Ciencia y Technologia (Spain).

\begin{table}
\squeezetable
\label{kin_tbl}
\caption{Kinematics for the present experiment. For the electron and
proton angles we indicate between parentheses the angles for the
$^1$H($\vec e,e^\prime\vec p\,$) reaction, if different from the
\hept{} reaction.}
\begin{tabular}{|c|c|c|c|c|c|}
  Beam   &  $Q^2$ & Electron & Electron       & Proton & Proton         \\
 Energy  &        & Momentum & $\theta_{LAB}$ & Momentum &$\theta_{LAB}$ \\
 (MeV)   & (GeV/c)$^2$ & (MeV/c) & (degrees)    & (MeV/c)  & (degrees)      \\
\hline
3400 & 0.5  & 3102 & 12.47(12.50) &  766 & 61.43(63.12) \\
4239 & 1.0  & 3667 & 14.56        & 1150 & 54.55(54.82) \\
4237 & 1.6  & 3340 & 19.35        & 1549 & 45.75(46.77) \\ 
4237 & 2.6  & 2796 & 27.10        & 2161 & 36.20 \\
\end{tabular}
\end{table}

\begin{table}
\squeezetable
\label{ratio_tbl}
\caption{Polarization ratios with statistical and estimated systematic
uncertainties. The polarization ratio value for
$^1$H($\vec e$,e$^\prime \vec p$) at $Q^2$ = 2.6 (GeV/c)$^2$ is from the
fit of Ref. \protect\cite{gegmprl}. The uncertainty in this ratio
and in $R$ reflects the typical systematic uncertainty of the data of 
Ref. \protect\cite{gegmprl} at this $Q^2$.}
\begin{tabular}{|c|c|c|c|}
$Q^2$ & ($P^\prime_x/P^\prime_z$)$_{He}$ & ($P^\prime_x/P^\prime_z$)$_{H}$ &
$R$ \\
\hline
0.5 & -0.804$\pm$0.035$\pm$0.006 & -0.898$\pm$0.029$\pm$0.011 & 0.895$\pm$0.048$\pm$0.015 \\
1.0 & -0.502$\pm$0.018$\pm$0.005 & -0.578$\pm$0.014$\pm$0.005 & 0.868$\pm$0.038$\pm$0.011 \\
1.6 & -0.393$\pm$0.014$\pm$0.011 & -0.395$\pm$0.010$\pm$0.009 & 0.992$\pm$0.043$\pm$0.007 \\
2.6 & -0.231$\pm$0.022$\pm$0.016 & (-0.265$\pm$0.024) & 0.869$\pm$0.081$\pm$0.099 \\
\end{tabular}
\end{table}

\begin{thebibliography}{99}
\bibitem{Dipc} With the initial and final electron momentum $\vec k_i$ and
$\vec k_f$, the coordinate system is given by the unit vectors
$\hat z = (\vec k_i - \vec k_f) / |\vec k_i - \vec k_f |$,
$\hat y = (\vec k_i \times \vec k_f) / |\vec k_i \times \vec k_f |$, and
$\hat x = \hat y \times \hat z$.
\end{thebibliography}
\end{document}